\documentstyle[floats,epsfig,twocolumn,prd,aps]{revtex}

\begin{document} \draft

\newcommand{\be}{\begin{equation}}
\newcommand{\ee}{\end{equation}}
\newcommand{\ba}{\begin{eqnarray}}
\newcommand{\ea}{\end{eqnarray}}
\newcommand{\nn}{\nonumber\\}
\newcommand{\n}[1]{\label{#1}}


\wideabs{
\title{Black Holes in a Compactified Spacetime}
\author{Andrei V. Frolov$^*$ and Valeri P. Frolov$^\dagger$}
\address{
  \medskip
  $^*$CITA, University of Toronto\\
  Toronto, ON, Canada, M5S 3H8\\
  {\rm E-mail: \texttt{frolov@cita.utoronto.ca}}
}
\address{
  \medskip
  $^\dagger$Theoretical Physics Institute, 
  Department of Physics, University of Alberta\\
  Edmonton, AB, Canada, T6G 2J1\\
  {\rm E-mail: \texttt{frolov@phys.ualberta.ca}}
  \medskip
}
\date{12 February 2003}
\maketitle

\begin{abstract}
  We discuss properties of a 4-dimensional Schwarzschild black hole in a
  spacetime where one of the spatial dimensions is compactified. As a
  result of the compactification the event horizon of the black hole is
  distorted. We use Weyl coordinates to obtain the solution describing
  such a distorted black hole. This solution is a special case of the
  Israel-Khan metric. We study the properties of the compactified
  Schwarzschild black hole, and develop an approximation which allows
  one to find the size, shape, surface gravity and other characteristics
  of the distorted horizon with a very high accuracy in a simple
  analytical form. We also discuss possible instabilities of a black
  hole in the compactified space.
\end{abstract}

\pacs{PACS numbers: 04.50.+h, 98.80.Cq \hfill CITA-2003-04, Alberta-Thy-03-03}
}
\narrowtext


\section{Introduction}\label{s1}

Black hole solutions in a compactified spacetime have been studied in
many publications. A lot of attention was paid to Kaluza-Klein
higher dimensional black holes. By compactifying black hole solutions
along Killing directions one obtains a lower dimensional solutions
of Einstein equations with additional scalar, vector and other fields
(see e.g. \cite{GiWi:86} and references therein). The generation of
black hole and black string solutions by the Kaluza-Klein procedure
was extensively used in the string theory (see e.g. \cite{Lars:00}
and references therein).

A solution which we consider in this paper is of a different nature. We
study a Schwarzschild black hole in a spacetime with one compactified
spatial dimension. This dimension does not coincide with any Killing
vector, for this reason the black hole metric is distorted as a result
of the compactification.

Recent interest in compactified spacetimes with black holes is
connected with brane-world models. General properties of black holes in
the Randall-Sundrum model were discussed in \cite{ChHaRe:00,EmHoMy:00}.
In the latter paper 4-dimensional C-metric was used to obtain an exact
3+1 dimensional black hole solution in AdS spacetime with the
Randall-Sundrum brane. Black holes in RS braneworlds were discussed in
a number of publications (see e.g.~\cite{KuTaNa:03} and references
therein).

Black hole solutions in a spacetime with compactified dimensions are
also interesting in connection with other type of brane models, which
were considered historically first in \cite{BW}. In ADD-type of brane
worlds the tension of the brane can be not very large. If one neglects
its action on the gravitational field of a black hole, one obtains a
black hole in a spacetime where some of the dimensions are
compactified. Compactification of special class of solutions,
generalized Majumdar-Papapetrou metrics, was discussed by Myers
\cite{Myers:87}. In this paper he also made some general remarks
concerning compactification of the 4-dimensional Schwarzschild metric.
Some of the properties of compactified 4-dimensional Schwarzschild
metrics were also considered in \cite{BoPe:90}. For a recent discussion
of higher dimensional black holes on cylinders see \cite{HaOb:02}.

In this paper we study a solution describing a 4-dimensional
Schwarzschild black hole in a spacetime where one of the dimensions is
compactified. This solution is a special case of the Israel-Khan metric
\cite{IsKh:64}. Its general properties were discussed by Korotkin and
Nicolai \cite{KoNi:94}. As a result of the compactification the event
horizon of the black hole is distorted. In our paper we focus our
attention on the properties of the distorted horizon. We use Weyl
coordinates to obtain a solution describing such a distorted black
hole. This approach to study of axisymmetric static black holes is well
known and was developed long time ago by Geroch and Hartle \cite{GeHa:81}
(see also \cite{FrNo:98})\footnote{For generalization of this approach
to the case of electrically charged distorted 4-D black holes see
\cite{FaKr:01,Yaza:01}. A generalization of the Weyl method to higher
dimensional spacetimes was discussed in \cite{EmRe:02}. An initial
value problem for 5-D black holes was discussed in \cite{ShSh:00,SoPi:02}}.
In Weyl coordinates, the metric describing a distorted 4-dimensional
black hole contains 2 arbitrary functions. One of them, playing a role
of the gravitational potential, obeys a homogeneous linear equation.
Because of the linearity, one can present the solution as a linear
superposition of the unperturbed Schwarzschild gravitational potential,
and its perturbation. After this, the second function which enters the
solution can be obtained by simple integration. To find the
gravitational potential one can either use the Green's function method
or to expand a solution into a series over the eigenmodes. We discuss
both of the methods since they give two different convenient
representations for the solution. We develop an approximation which
allows one to find the size, shape, surface gravity and other
characteristics of the distorted horizon with very high accuracy in a
simple analytical form. We study properties of compactified
Schwarzschild black holes and discuss their possible instability.

The paper is organized as follows. We recall the main properties of 4-D
distorted black holes in Section~2. In Section~3, we obtain the
solution for a static 4-dimensional black hole in a spacetime with 1
compactified dimension. In Section~4, we study this solution. In
particular we discuss its asymptotic form at large distances, and the
size, form and shape of the horizon. We conclude the paper by general
remarks in Section~5.


\section{4-dimensional Weyl black holes}

\subsection{Weyl form of the Schwarzschild metric}

A static axisymmetric 4 dimensional metric in the canonical Weyl
coordinates takes the form \cite{GeHa:81,EmRe:02,FrNo:98}
\be\n{2.1}
dS^2=-e^{2U}\, dT^2 +e^{-2U}\, \left[ e^{2V}\, (dR^2 +dZ^2)+
R^2\, d\phi^2 \right]\, ,
\ee
where $U$ and $V$ are functions of $R$ and $Z$. This metric is a
solution of vacuum Einstein equations if and only if these functions
obey the equations
\be\n{2.2}
{\partial^2 U\over \partial R^2}+{1\over R}\,{\partial U\over \partial
R}+
{\partial^2 U\over \partial Z^2} =0\, ,
\ee
\be\n{2.3}
V_{,R}=R\, (U_{,R}^2-U_{,Z}^2)\, ,\hspace{1em}
V_{,Z}=2R\, U_{,R}\, U_{,Z}\, .
\ee
Let
\be\n{2.4}
dl^2=dR^2 +R^2\, d\psi^2+dZ^2\,
\ee
be an auxiliary 3 dimensional flat metric, then solutions of
(\ref{2.2}) coincide with axially symmetric solutions of the 3
dimensional Laplace equation
\be\n{2.5}
\Delta\, U=0\, ,
\ee
where $\Delta$ is a flat Laplace operator in the metric (\ref{2.4}).
It is easy to check that the equation (\ref{2.2}) plays the role of
the integrability condition for the linear first order equations
(\ref{2.3}). The regularity condition implies that at regular points
of the symmetry axis $R=0$
\be\n{2.6}
\lim_{R\to 0}\, V(R,Z)=0\, .
\ee
In fact, if $V(0,Z_0)=0$ at any point $Z_0$ of $Z$-axis then
(\ref{2.3}) implies that $V(0,Z)=0$ at any other point of the $Z$-axis
which is connected with $Z_0$.

For a four dimensional Schwarzschild metric, the function $U$ is the
potential of an infinitely thin finite rod of mass $1/2$ per unit
length located at $-M\le Z\le M$ portion of the $Z$-axis
\be\n{2.7}
\Delta\, U=4\pi j\, ,\hspace{1em} j={1\over 4\pi}{\delta(\rho)\over
\rho}\, \Theta(z/M)\, ,
\ee
where
\be\n{2.8}
\Theta(x)=\left\{ \begin{array}{cc}
1\, , & \ \ |x|\le 1 \, ,\\
0 \, ,& \ \ |x|>1 \, .
\end{array}
\right.
\ee
The corresponding solution is
\ba\n{2.9}
U_S(R,Z)
  &=& -{1\over 2}\int\limits_{-M}^{M}\, {dZ'\over \sqrt{R^2+(Z-Z')^2}}\, \\
  &=& -{1\over 2}\, \ln\left[ {\sqrt{(M-Z)^2+R^2}-Z +M \over \sqrt{(M+Z)^2+R^2}-Z-M)} \right]\, .\nonumber
\ea
The integral representation in the right hand side of equation
(\ref{2.9}) is obtained by using the 3-dimensional Green's function
for the equation (\ref{2.5}), which is of the form
\ba\n{2.10}
G^{(3)}({\bf x},{\bf x}')
  &=& {1\over 4\pi |{\bf x}-{\bf x}'|}\,\\
  && \hspace{-4em}
   = {1\over 4\pi}\, {1\over\sqrt{R^2+{R'}^2-2RR'\cos(\psi-\psi')+(Z-Z')^2}}\, .\nonumber
\ea
Sometimes the solution (\ref{2.9}) is presented in another equivalent
form
\be\n{2.11}
U_S(R,Z)={1\over 2}\ln \left({L-M\over L+M}\right)\, ,
\ee
\be\n{2.12}
L={1\over 2}(L_+ + L_-)\, ,\hspace{1em}
L_{\pm}=\sqrt{R^2+(Z\pm M)^2}\, .
\ee
The function $V_S(R,Z)$ for the Schwarzschild metric can be found
either by solving equations (\ref{2.3}) or by direct change of the
coordinates
\be\n{2.13}
R=\sqrt{r(r-2M)}\sin\theta\, ,\hspace{1em}
Z=(r-M)\cos\theta\, .
\ee
One has
\be\n{2.14}
V_S(R,Z)={1\over 2}\ln \left({L^2-M^2\over L^2-\eta^2}\right)\,,
\ee
\be
\eta={1\over 2}(L_+ -L_-)\, .
\ee
In the coordinates $(R,Z)$ the black hole horizon $H$ is the line
segment $-M\le Z\le M$ of the $R=0$ axis.

\subsection{A distorted black hole}

General static axisymmetric distorted black holes were studied in
\cite{GeHa:81}. A distorted black hole is described by a static
axisymmetric Weyl metric with a regular Killing horizon. One can
write the solution $(U,V)$ for a distorted black hole as
\be\n{2.15}
U=U_S+\hat{U}\, ,\hspace{1em} V=V_S+\hat{V}\, ,
\ee
where $(U_S,V_S)$ is the Schwarzschild solution with mass $M$. Since
both $V$ and $V_S$ vanish at the axis $R=0$ outside the horizon,
the function $\hat{V}$ has the same property. The function $\hat{U}$
obeys the homogeneous equation (\ref{2.2}), while the equations for
$\hat{V}$ follow from (\ref{2.3}). One of these equations is of the
form
\be\n{2.16}
\hat{V}_{,Z}=2R\left(
{U_S}_{,R}\,\hat{U}_{,Z} + {U_S}_{,Z}\,\hat{U}_{,R}+\hat{U}_{,R}\,\hat{U}_{,Z} \right)\, .
\ee
Near the horizon $\hat{U}$ is regular, while ${U_S}_{,R}=O( R^{-1})$
and ${U_S}_{,Z}=O(1)$. Thus near the horizon $\hat{V}_{,Z}\sim
2\hat{U}_{,Z}$. Integrating this relation along the horizon from $Z=-M$
to $Z=M$ and using the relations $\hat{V}(0,-M)=\hat{V}(0,M)=0$, we
obtain that $\hat{U}$ has the same value $u$ at both ends of the line
segment $H$. By integrating the same equation along the segment $H$
from the end point to an arbitrary point of $H$ one obtains for $-M\le
Z\le M$
\be\n{2.17}
\hat{V}(0,Z)=2 \left[\hat{U}(0,Z)-u\right]\, .
\ee
Geroch and Hartle \cite{GeHa:81} demonstrated that if $\hat{U}$ is a
regular smooth solution of (\ref{2.5}) in any small open neighborhood
of $H$ (including $H$ itself) which takes the same values, $u$, on
the both ends of the segment $H$, then the solution is regular at the
horizon and describes a distorted black hole.

Using the coordinate transformation
\ba\n{2.18}
R&=&e^u\, \sqrt{r(r-2M_0)}\, \sin\theta\, ,\\
Z&=&e^u\, (r-M_0)\, \cos\theta\, ,\nonumber
\ea
and defining
\be\n{2.19}
M_0=M\, e^{-u}\,,
\ee
it is possible to recast the metric (\ref{2.1}) of a distorted black
hole into the form
\ba\n{2.20}
dS^2 &=& -e^{-2\hat{U}}\, \left(1-{2M_0\over r}\right)dT^2 \\
     && +e^{2(\hat{V}-\hat{U}+u)}\, \left(1-{2M_0\over r}\right)^{-1}\, dr^2 \nn
     && +e^{2(\hat{V}-\hat{U}+u)}\, r^2\, \left(d\theta^2+e^{-2\hat{V}}\, \sin^2\theta\, d\phi^2\right)\, .\nonumber
\ea
In these coordinates, the event horizon is described by the equation
$r=2M_0$, and the 2-dimensional metric on its surface is
\be\n{2.21}
d\gamma^2
=4M_0^2\, \left[ e^{2(\hat{U}-u)}\, d\theta^2
+e^{-2(\hat{U}-u)}\, \sin^2\theta\, d\phi^2 \right]\, .
\ee
The horizon surface has area
\be\n{2.22}
A=16\pi M_0^2\, .
\ee
It is a sphere deformed in an axisymmetric manner. The surface gravity
$\kappa$ is constant over the horizon surface:
\be\n{2.23}
\kappa={e^u\over 4M_0}\, .
\ee

\section{4-D compactified Schwarzschild black hole}

\subsection{Compactified Weyl metric}

In what follows it is convenient to rewrite the Weyl metric
(\ref{2.1}) in the dimensionless form $dS^2=L^2\, ds^2$,
\be\n{3.1}
ds^2=-e^{2U}\, dt^2 +e^{-2U}\, \left[ e^{2V}\, (d\rho^2 +dz^2)+
\rho^2\, d\phi^2 \right]\, ,
\ee
where $L$ is the scale parameter of the dimensionality of the length
and
\be\n{3.2}
t={T\over L}\, ,\quad \rho={R\over L}\, ,\quad z={Z\over L}\, .
\ee
are dimensionless coordinates. We shall also use instead of mass $M$
its dimensionless version $\mu=M/L$. The Schwarzschild solution
(\ref{2.9}) then can be rewritten as
\be\n{3.3}
U_S(\rho,z)= -{1\over 2}\, \log\left[ {\sqrt{
(\mu-z)^2+\rho^2}-z +\mu \over \sqrt{
(\mu+z)^2+\rho^2}-z-\mu)} \right]\, .
\ee

For $|z|>\mu$, the gravitational potential $U_S$ remains finite at the
symmetry axis
\be\n{3.4}
U_S(0,z)= {1\over 2}\,\ln {z-\mu\over z+\mu}\, ,\hspace{0.5cm}
|z|>\mu\, .
\ee
For $|z|\le \mu$, the gravitational potential $U_S$ is divergent at
$\rho=0$. The leading divergent term is
\be\n{3.5}
U_S(\rho,z)\sim {1\over 2}\,\ln {\rho^2\over 4(\mu^2-z^2)}\, ,\hspace{0.5cm}
|z|\le \mu\, .
\ee

We will now obtain a new solution describing a Schwarz\-schild black
hole in a space in which $Z$-coordinate is compactified. We will call
this solution a compactified Schwarzschild metric, or briefly
CS-metric. For this purpose we assume that the coordinate $Z$ is
periodic with a period $2\pi L$. We shall use the radius of
compactification $L$ as the scale factor.

Our space manifold ${\cal M}$ has topology $S^1\times R^2$ and we are
looking for a solution of the equation (\ref{2.5}) on ${\cal M}$ which
is periodic in $z$ with the period $2\pi$, \ $z\in (-\pi,\pi)$. The
source for this solution is an infinitely thin rod of the linear
density $1/2$ located along $z$ axis in the interval $(-\mu,\mu)$, \,
$\mu\le \pi$. This problem can be solved by two different methods,
either by using Green's functions or by expanding a solution into a
series over the eigenmodes. We discuss both of the methods since they
give two different convenient representations for the solution. We
begin with the method of Green's functions.

\subsection{3-D Green's function}

To obtain this solution we proceed as follows. Our first step is to
obtain a 3-dimensional Green's function $G^{(3)}_{\cal M}$ on the
manifold ${\cal M}$. It can be done, for example, by the method of
images applied to the Green's function for equation (\ref{2.5}) which
gives the series representation for $G^{(3)}_{\cal M}$. It is more
convenient to use another method which gives the integral
representation. For this purpose we note that the flat 3-dimensional
Green's function can be obtained by the dimensional reduction from the
4-dimensional one. Namely let ${\bf X}=(X,Y,Z,W)$
\be\n{3.6}
dh^2=d{\bf X}^2=dX^2+dY^2+dZ^2+dW^2\, ,
\ee
then
\be\n{3.7}
G^{(3)}({\bf x},{\bf x}')\equiv {1\over 4\pi |{\bf x}-{\bf x}'|}
=\int\limits_{-\infty}^{\infty} \, dW \, G^{(4)}({\bf X},{\bf X}')\, ,
\ee
where ${\bf x}=(X,Y,Z)$,
\be\n{3.8}
G^{(4)}({\bf X},{\bf X}')={1\over 4\pi^2}{1\over |{\bf X}-{\bf X}'|^2}\, ,
\ee
and $G^{(4)}(X,X')$ is the Green's function for the Laplace
operator
\be\n{3.9}
\Delta^{(4)}\, G^{(4)}({\bf X},{\bf X}')=-\delta^4({\bf X}-{\bf X}')\, ,
\ee

Denote
\be\n{3.10}
G^{(4)}_{\cal M}({\bf X},{\bf X}')= {1\over 4\pi^2}\, \sum_{n=-\infty}^{\infty}\,
{1\over (Z-Z'+2\pi L n)^2+ B^2}\, .
\ee
where
\be\n{3.11}
B^2=(X-X')^2+(Y-Y')^2+(W-W')^2\, .
\ee
The function $G^{(4)}_{\cal M}$ is periodic in $Z$ with the period
$2\pi L$ and is a Green's function on the manifold ${\cal M}$. The sum
can be calculated explicitly by using the following relation
\be\n{3.12}
\sum_{-\infty}^{\infty}{1\over (a+n)^2+b^2}={\pi\over b}{\sinh(2\pi
b)\over \cosh(2\pi b)-\cos(2\pi a)}\, .
\ee
Thus one has
\be\n{3.13}
G^{(4)}_{\cal M}({\bf X},{\bf X}')= {1\over 8\pi^2 L^2 \beta}\,
{\sinh \beta\over [\cosh\beta-\cos(z-z')]}\, ,
\ee
where $\beta =B/L$. This Green's function has a pole at $\beta=z-z'=0$,
that is when the points ${\bf X}$ and ${\bf X}'$ coincide. At far
distance, $\beta\gg L$, this Green's function has asymptotic
\be\n{3.14}
G^{(4)}_{\cal M}({\bf X},{\bf X}')\sim {1\over 8\pi^2 L^2 \beta}\, ,
\ee
and hence it behaves as if the space had one dimension less. It
is obviously a result of compactification.

In the reduction procedure this creates a technical problem since the
integral over $w$ becomes divergent. It is easy to deal with this
problem as follows. Denote
\be\n{3.15}
G^{(4,\alpha)}_{\cal M}({\bf X},{\bf X}')= G^{(4,\text{reg})}_{\cal M}({\bf X},{\bf X}')
+{1\over 8\pi^2 L^2 }{1\over (\beta^2+b^2)^{\alpha/2}}\, ,
\ee
\ba\n{3.16}
G^{(4,\text{reg})}_{\cal M}({\bf X},{\bf X}') = \hspace{-6em} &&\\
&& = {1\over 8\pi^2 L^2 } \left[{1\over \beta}\,
\,{\sinh \beta\over \cosh\beta-\cos(z-z')}
-{1\over \sqrt{\beta^2+b^2}}\right]\, . \nonumber
\ea
Here $b$ is any positive number. For $\alpha=1$, $G^{(4,\alpha)}_{\cal
M}$ does not depend on $b$ and coincides with (\ref{3.7}). At large
$\beta$ the term $G^{(4,\text{reg})}_{\cal M}$ has asymptotic behavior
$\sim \beta^{-2}$.

We also have
\ba\n{3.17}
\int\limits_{-\infty}^{\infty}\, {dw\over (\beta^2+b^2)^{\alpha/2}} = \hspace{-7em} && \\
 && = {1\over 2\sqrt{\pi}}{ [\sigma^2+b^2]^{(1-\alpha)/2} \Gamma( (\alpha-1)/2)\over \Gamma(\alpha/2)}\, \nn
 && \sim {1\over \pi}\, \left[ {1\over \alpha-1}+\ln 2 -{1\over 2}\ln (\sigma^2+b^2) \right] +O(\alpha-1) \nonumber\, .
\ea
Here $\sigma^2=(x-x')^2+(y-y')^2$. By omitting unimportant (divergent)
constant we regularize the expression for the integral.

By using the reduction procedure (\ref{3.7}) we get
\ba\n{3.18}
G^{(3)}_{\cal M}({\bf x},{\bf x}') &=& \int\limits_{-\infty}^{\infty}\, dW\,
G^{(4,\text{reg})}_{\cal M}({\bf X},{\bf X}') \\
 &&\hspace{1.5em} -{1\over 16\pi^2 L}\, \ln (\sigma^2+b^2)\, . \nonumber
\ea

\subsection{Integral representation for the gravitational potential}

To obtain the potential $U(\rho,z)$ which determines the black hole
metric we need to integrate $G^{(3)}_{\cal M}({\bf x},{\bf x}')$ with
respect to ${\bf x}'$ along the interval $(-M,M)$ at $R'=0$ axis. It is
convenient to use the representation (\ref{3.18}) and to change the
order of integrals. We use the following integral $(a>1, 0<\mu<\pi,
-\pi< z <\pi)$
\ba\n{3.19}
  \int\limits_{-\mu}^{\mu}\, {dz'\over a-\cos(z'-z)} = \hspace{-8em} &&\\
  && = {2\over \sqrt{a^2-1}}\left\{\textstyle
    \arctan\left[p\tan\left({\mu+z\over 2}\right)\right]
    +\pi\vartheta(\mu+z-\pi) \right.\nn
  &&\left.\textstyle \hspace{4.5em} +
    \arctan\left[p\tan\left({\mu-z\over 2}\right)\right]
    +\pi\vartheta(\mu-z-\pi)\right\} \, ,\nonumber
\ea
where $p=\sqrt{a+1\over a-1}$. We understand $\arctan$ to be the
principal value and include $\vartheta$-functions to get the correct
value over all the interval $-\pi <z< \pi$. We also change the
parameter of integration $W$ to $w=W/L$ and take into account that the
integrand is an even function of $w$. After these manipulations we
obtain
\ba\n{3.20}
U(\rho,z) &=& -{1\over \pi}
\int\limits_0^{\infty}\, dw\left(
{{\cal U}(\beta,z)\over \beta}-{\mu\over
\sqrt{\beta^2+b^2}}
\right)\\
 && \hspace{4em} +{\mu\over 2\pi}\ln(\rho^2+b^2)\, ,\nonumber
\ea
where
\ba\n{3.21}
{\cal U}(\beta,z)&=&{\cal V}(\beta,z) + {\cal V}(\beta,-z)\, ,\\
\n{3.22}
{\cal V}(\beta,z)&=&\textstyle
\arctan\left[
{\cosh\beta+1\over \sinh\beta}\,
\tan\left({\mu+z\over 2}\right)
\right]+\pi\vartheta(\mu+z-\pi)\, .\hspace{-1em}\nonumber
\ea
Note that now $\beta$ which enters equations (\ref{3.20}) and
(\ref{3.21}) is
\be\n{3.23}
\beta=\sqrt{w^2+\rho^2}\, .
\ee

A representation similar to (\ref{3.20}) can be written for the
Schwarzschild potential $U_S$
\be\n{3.24}
U_S(\rho,z)=-{1\over \pi}
\int\limits_0^{\infty}\, dw\ 
{{\cal U}_S(\beta,z)\over \beta}\, ,
\ee
where
\ba\n{3.25}
{\cal U}_S(\beta,z)&=&{\cal V}_S(\beta,z)
 +{\cal V}_S(\beta,-z)\, ,\\
{\cal V}_S(\beta,z)&=&\textstyle
\arctan\left({\mu+z\over \beta}\right)\, .\nonumber
\ea
One can check that this integral really gives expression (\ref{3.3}).

Using these representations we obtain the following expression for the
quantity $\hat{U}(z)=U(0,z)-U_S(0,z)$ which determines the properties of
the event horizon
\be\n{3.26}
\hat{U}(z)=-{1\over \pi}
\int\limits_0^{\infty}\, dw\left(
{{\cal U}(w,z)-{\cal U}_S(w,z)
\over w}-{\mu\over \sqrt{w^2+1}}
\right)\, .
\ee
To obtain the redshift factor $u$ it is sufficient to calculate
$\hat{U}(z)$ for $z=\mu$
\be\n{3.27}
u=\hat{U}(\mu)\, .
\ee

\subsection{Series representation for the gravitational potential}

For numerical calculations of the gravitational potential $U$ and study
of its asymptotics near the black hole horizon it is convenient to use
another representation for $U$, namely its Fourier decomposition with
respect to the periodic variable $z$. Note that a function
$\Theta(z/\mu)$ which enters the source term (see equations
(\ref{2.7}--\ref{2.8})) allows the following Fourier decomposition
on the circle
\be\n{3.28}
\Theta(z/\mu)=a_0+\sum_{k=1}^{\infty}\, a_k\ \cos(kz)\, ,
\ee
where
\be\n{3.29}
a_0={\mu\over \pi}\, ,\hspace{0.5cm} a_k={2\over \pi k}\sin(k \mu)\, .
\ee
Using the Fourier decomposition for $U$
\be\n{3.30}
U(\rho,z)=U_0(\rho)+\sum_{k=1}^{\infty} U_k(\rho)\, \cos(kz)\, ,
\ee
we obtain the following equations for the radial functions $U_k(\rho)$
\be\n{3.31}
{d^2 U_k\over d\rho^2}+{1\over \rho}\, {d U_k\over d\rho} -k^2\, U_k=
a_k\, {\delta(\rho)\over \rho}\, .
\ee
For $k>0$ the solutions of these equations which are decreasing at
infinity are
\be\n{3.32}
U_k(\rho)=-a_k K_0(k\rho)\, ,
\ee
where $K_{\nu}(z)$ is MacDonald function. For $k=0$ the solution is
\be\n{3.33}
U_0(\rho)=a_0 \ln(\rho)\, .
\ee
Thus the gravitational potential $U$ allows the following series
representation
\be\n{3.34}
U(\rho, z)={\mu\over \pi}\, \ln \rho
-2\sum_{k=1}^{\infty}\, {\sin(k\mu)\over \pi k}\, \cos(kz)\, K_0(k\rho)
\, .
\ee

This representation is very convenient for studying the asymptotics
of the gravitational potential near the horizon. For small $\rho$ one
has
\be\n{3.35}
-K_0(k\rho)\sim \ln\left({k\rho\over 2} \right)+\gamma \, ,
\ee
where $\gamma \approx 0.57721$ is Euler's constant. Substituting these
asymptotics into (\ref{3.34}) and combining the terms one obtains
\ba\n{3.36}
U(\rho,z) &\sim&
  \left[ \ln{\rho\over 2}\,+\gamma \right]\, \Theta(z/\mu)
  -{\mu\over \pi}\,\left(\ln {1\over 2}\,+\gamma \right)\, \\
  &&
  +{1\over \pi}\, \sum_{k=1}^{\infty}\, {\ln k\over k}\,
  \Big[\sin(k(\mu+z))+\sin(k(\mu-z))\Big]\, .\nonumber
\ea
Using the relation (see equation 5.5.1.24 in \cite{PBM})
\ba\n{3.37}
\sum_{k=1}^{\infty}\, {\ln k\over k}\,\sin(kx) &=&
  {x-\pi\over 2}\, (\gamma+\ln 2\pi) \\
  &&\hspace{-2em} + {\pi\over 2}\, \ln \left| {1\over \pi}\sin {x \over 2}\,
    \Gamma^2\left({x\over 2\pi}\right) \right| \nonumber
\ea
valid for $0 \le x < 2\pi$, one gets for $|z|\le \mu$
\ba\n{3.38}
U(\rho,z) &\sim& \ln {\rho\over\pi}\, + {\mu\over\pi}\, \ln(4\pi) \\
  && +{1\over 2}\ln \left|{1\over \pi}\, \sin{\mu+z\over 2}\,
    \Gamma^2\left({\mu+z\over 2\pi}\right) \right| \nn
  && +{1\over 2}\ln \left|{1\over \pi}\, \sin{\mu-z\over 2}\,
    \Gamma^2\left({\mu-z\over 2\pi}\right) \right|\, .\nonumber
\ea
Using asymptotic (\ref{3.5}) of the Schwarzschild potential $U_S$ near horizon,
one can present $\hat{U}(z)=\lim\limits_{\rho\to 0}[U(\rho,z)-U_S(\rho,z)]$
in the region $|z|\le \mu$ in the following form
\be\n{3.39}
\hat{U}(z) = {\mu\over \pi}\, \ln(4\pi) + {1\over 2}\, \ln \left[\textstyle
f({\mu+z\over 2})f({\mu-z\over 2})\right]\, ,
\ee
where the function $f(x)$ is defined by
\be\n{3.40}
f(x)={1\over \pi^2}\, x\, \sin x\, \Gamma^2(\textstyle{x\over\pi})\, .
\ee
It has the following properties:
\be\n{3.41}
f(0)=1\, ,\hspace{1em}f(\textstyle{\pi \over 2})={1 \over 2} \, ,\hspace{1em}f(\pi)=0\, .
\ee
In fact, in the interval $0\le x \le \pi$ it can be approximated by a
linear function
\be\n{3.42}
f(x)\approx 1-{x\over \pi}\,
\ee
with an accuracy of order of 1\% .

Making similar calculations for $|z|\ge \mu$ one obtains
\be\n{3.38a}
U(0,z)= {\mu\over \pi}\, \ln (4\pi)
+{1\over 2}\ln \left[ { f({|z|+\mu\over 2})\over f({|z|-\mu\over
2})}\right]+{1\over 2}\ln {|z|-\mu \over |z|+\mu}\, .
\ee
An approximate value of $U(0,z)$ in this region is
\be\n{3.38b}
U(0,z)\approx {\mu\over \pi}\, \ln (4\pi)
+{1\over 2}\ln \left[ {[2\pi-(|z|+\mu)](|z|-\mu)\over
[2\pi-(|z|-\mu)](|z|+\mu)}\right]\, .
\ee

\subsection{Solutions}

\begin{figure*}
\begin{center}\begin{tabular}{c@{\hspace{1cm}}c}
  \epsfig{file=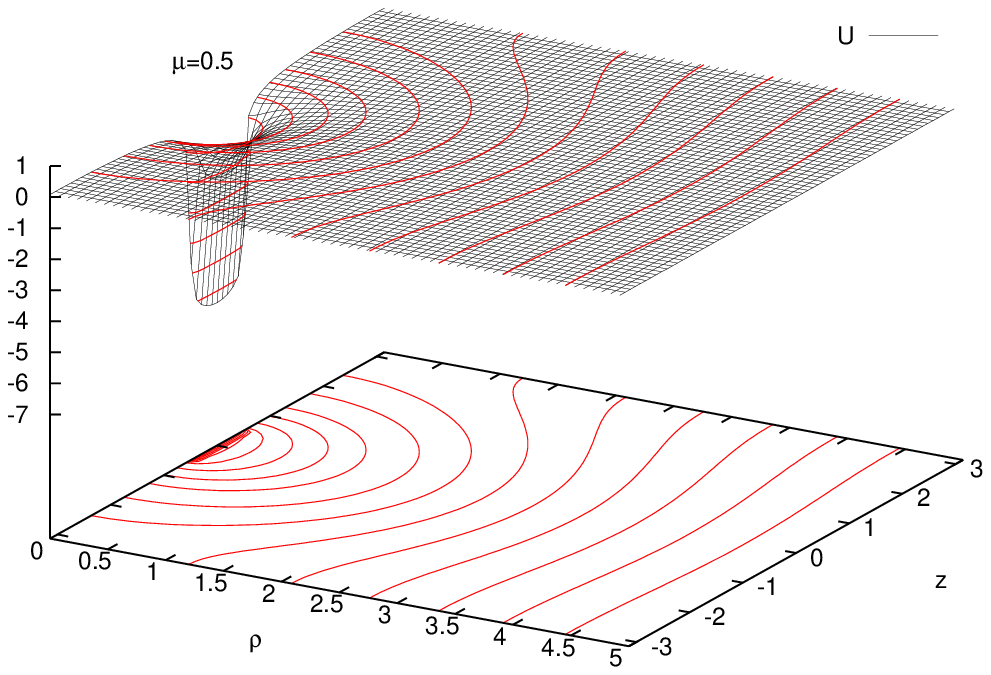, width=8.5cm} &
  \epsfig{file=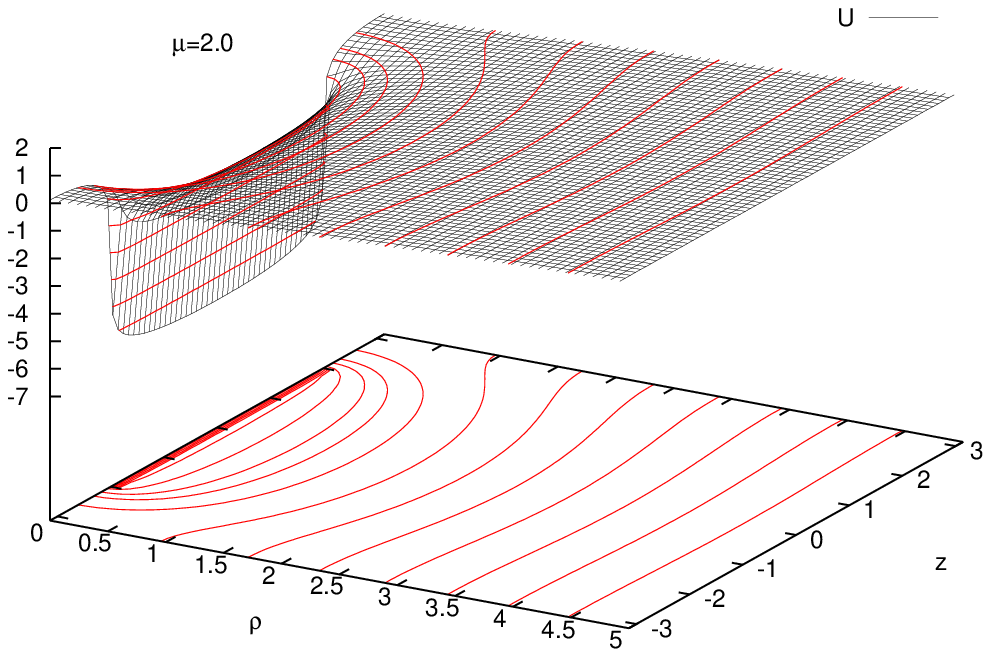, width=8.5cm} \\
  \\ \\
  \epsfig{file=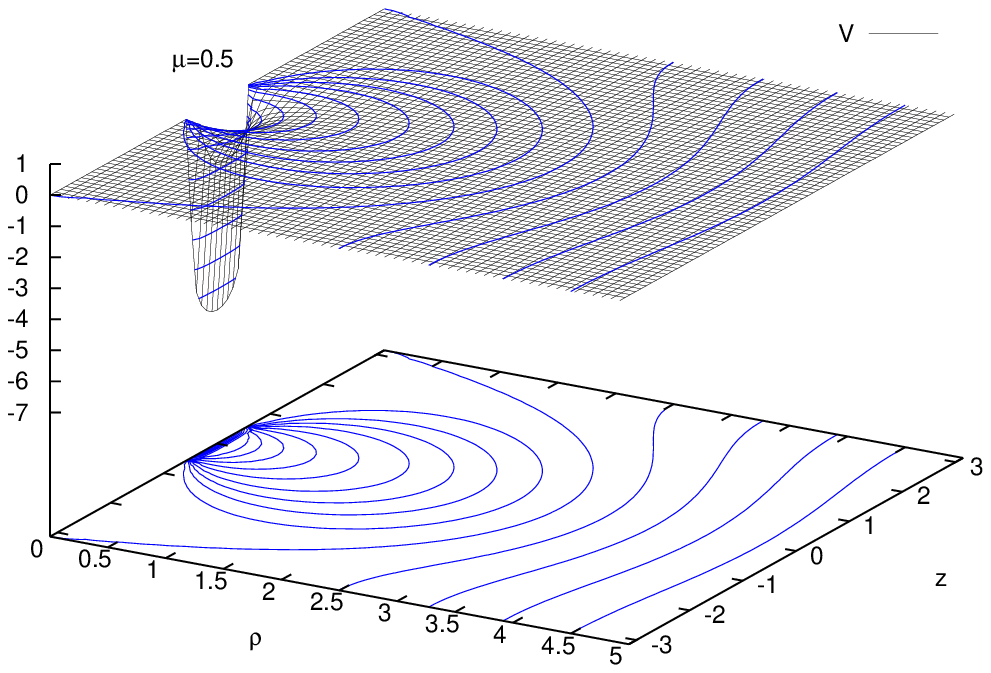, width=8.5cm} &
  \epsfig{file=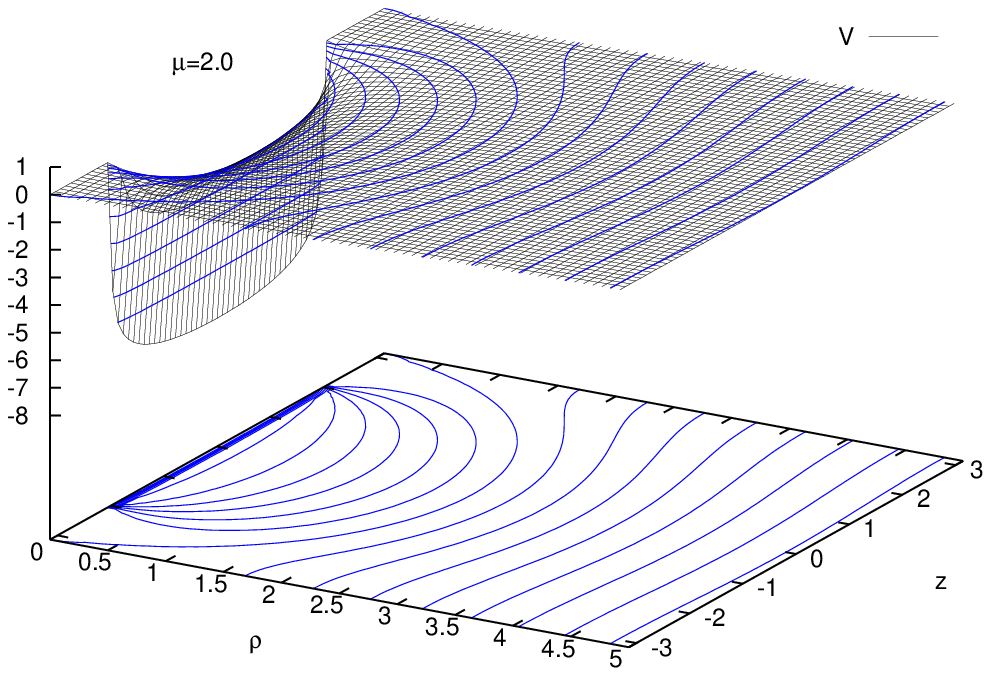, width=8.5cm} \\
\end{tabular}\end{center}
\caption{
  Compactified Schwarzschild black hole solutions for $\mu=0.5$ (left)
  and $\mu=2.0$ (right). The surface plots show the gravitational
  potential $U(\rho,z)$ (top) and the function $V(\rho,z)$ (bottom);
  red and blue contours represent equipotential surfaces of $U$ and $V$
  correspondingly.
}
\label{equipot}
\end{figure*}

To find the gravitational potential $U(\rho,z)$ one can use either its
integral representation (\ref{3.20}) or the series (\ref{3.34}). We
used both methods. Integrals (\ref{3.20}) were evaluated using Maple,
while the series (\ref{3.34}) were implemented in C code using FFT
techniques. Both methods give results which agree with high accuracy,
but of course the C implementation is much more computationally
efficient. The function $V(\rho,z)$ was recovered by direct integration
of differential equation (\ref{2.3}) by finite differencing in
$Z$-direction. The gravitational potential $U(\rho,z)$, function
$V(\rho,z)$, and their equipotential surfaces for two different values
of $\mu$ are shown in Figure~\ref{equipot}.

\section{Properties of CS black holes}

\begin{figure*}
  \centerline{
    \epsfig{file=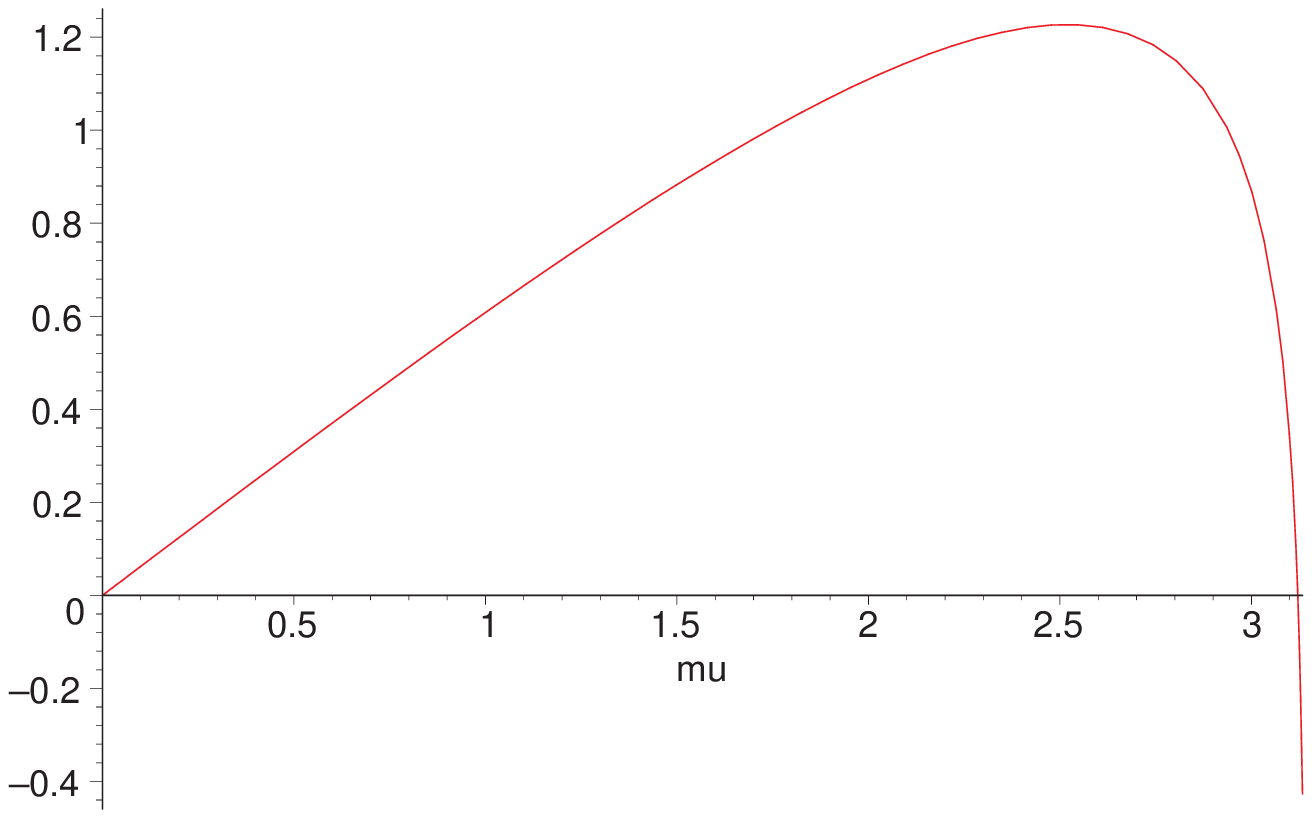, width=3in}\hspace{1cm}
    \epsfig{file=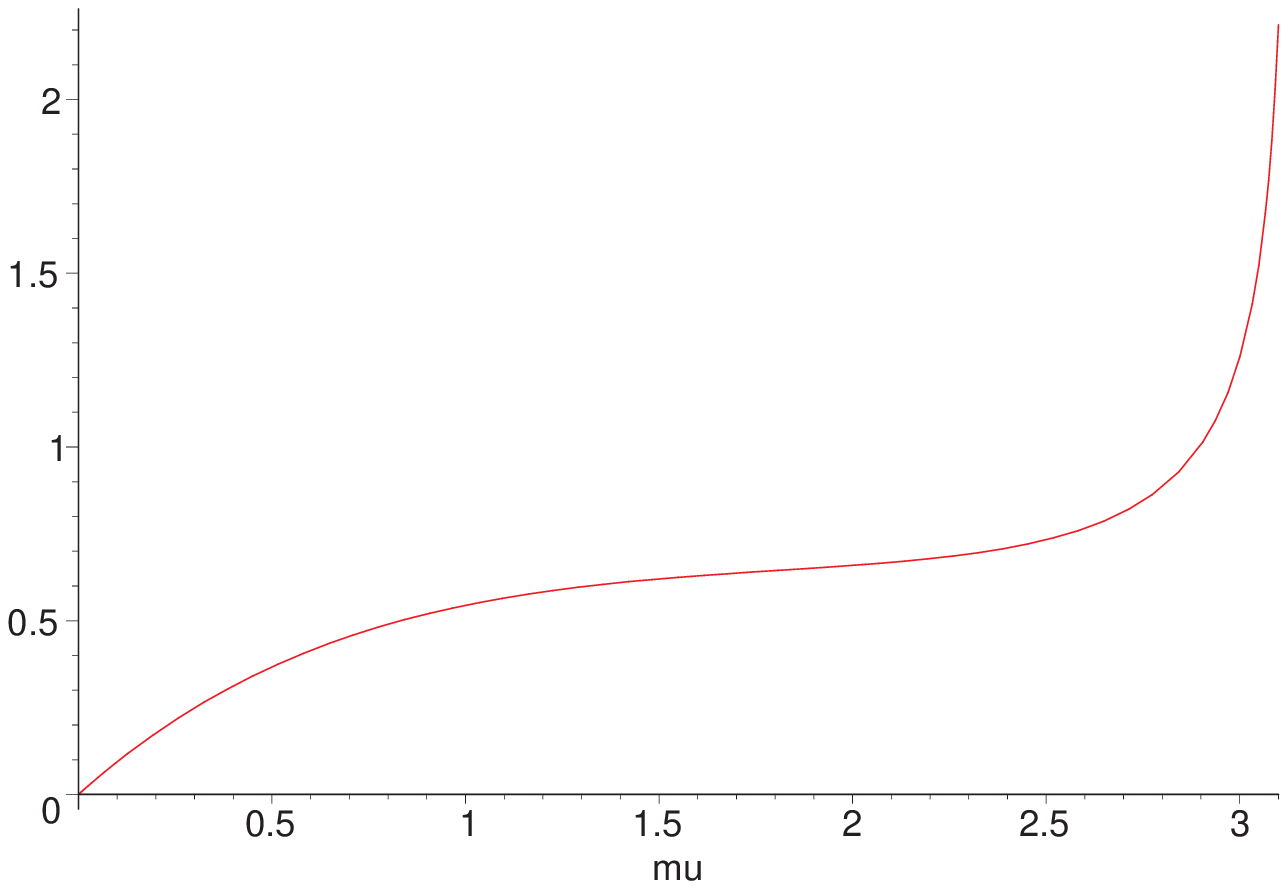, width=3in}
  }
  \caption{
    Redshift factor $u$ (left) and the irreducible mass $\mu_0=\mu\, \exp(-u)$ as functions of $\mu$.
  }
  \label{u}
\end{figure*}

\subsection{Large distance asymptotics}

Let us first analyze the asymptotic behavior of the CS-metric at large
distance $\rho$. For this purpose we use the integral representation
(\ref{3.20}) for $U$. It is easy to check that the integrand
expression at large $\rho$ is of order of $O(\beta^{-2})$ and hence
the integral is of order of $\rho^{-1}$. Thus the ln-term in the square
brackets in (\ref{3.20}) is leading at the infinity so that
\be\n{4.1}
U(\rho,z)|_{\rho\to\infty}\sim {\mu\over \pi}\ln\rho\, .
\ee\n{4.2}
Using equation (\ref{2.3}) we also get
\be\n{4.3}
V(\rho,z)|_{\rho\to\infty}\sim {\mu^2\over \pi^2}\ln\rho\, .
\ee
The metric (\ref{3.1}) in the asymptotic region $\rho\to\infty$ is of the form
\be\n{4.4}
ds^2=-\rho^{2{\mu\over\pi}}\, dt^2+\rho^{-2{\mu\over\pi}(1-{\mu\over\pi})}\,
(d\rho^2+dz^2)+ \rho^{-2{\mu\over\pi}}\, \rho^2\, d\phi^2 \, .
\ee
The proper size of a closed Killing trajectory for the vector
$\partial_z$ is
\be\n{4.4a}
C_z=2\pi L\rho^{-{\mu\over \pi}(1-{\mu\over \pi})}\, .
\ee

The metric (\ref{4.4}) coincides with the special case $(a_1=a_2)$ of
the Kasner solution \cite{Kasner}
\ba\n{4.5}
&ds^2=-\rho^{2a_0}\, dt^2+ \rho^{2a_1}\, d\rho^2+ \rho^{2a_2}\, dz^2
+\rho^{2a_3}\, d\phi^2\, ,&\\
\n{4.6}
&a_1+1=a_2+a_3+a_0\,,&\nn
&(a_1+1)^2=a_2^2+a_3^2+a_0^2\, .&\nonumber
\ea

One can rewrite the metric (\ref{4.3}) by using the proper-distance
coordinate $l$. For small $\mu$
\be\n{4.7}
l={\rho^{1-{\mu\over\pi}}\over 1-{\mu\over\pi}}\, ,
\ee
and the metric in the $(\rho,\phi)$-sector takes the form
\be\n{4.8}
dl^2+\left(1-{\mu\over\pi}\right)^2\, l^2\, d\phi^2\, .
\ee
Thus the metric of the CS black hole has an angle deficit $2\mu$ at
infinity.

The asymptotic form of the metric can be used to determine the mass of
the system. Let $\xi_{(t)}^{\mu}$ be a timelike Killing vector and
$\Sigma$ be a 2-D surface lying inside $t$=const hypersurface, then
the Komar mass $m$ is defined as
\be\n{4.8a}
m={1\over 4\pi}\, \int_{\Sigma}\, \xi_{(t)}^{\mu;\nu}\, d\sigma_{\mu\nu}\, .
\ee
For simplicity we choose $\Sigma$ so that $t$=const and
$\rho=\rho_0$=const. For this choice
\ba\n{4.8b}
d\sigma_{\mu\nu} &=& {1\over 2}\, \delta^{0}_{[\mu}\,
\delta^{1}_{\nu]}\,\rho_0^{1+2a_1}\, dz\, d\phi\, ,\\
\n{4.8c}
\xi_{\mu;\nu} &=& -2a_0\, \rho_0^{2a_0-1}\, \delta_{[\mu}^{0}\,
\delta_{\nu]}^{1}\, ,\nn
\xi^{\mu;\nu} &=& 2a_0\, \rho_0^{-2a_1-1}\, \delta^{[\mu}_{0}\,
\delta^{\nu]}_{1}\, .\nonumber
\ea
Substituting these expressions into (\ref{4.8a}) and taking the
integral we get $m=\mu$. Since all our quantities are normalized by the
radius of compactification $L$, we obtain that the Komar mass of our
system is $M=L\mu$.

\subsection{Redshift factor, surface gravity, and proper distance
between black hole poles}

Using equation (\ref{3.39}), we obtain for the redshift factor $u$ the
following expression
\be\n{3.43}
u={\mu\over \pi}\ln (4\pi)+{1\over 2}\ln f(\mu)\, .
\ee
Figure~\ref{u}(left) shows dependence of the redshift factor $u$ on
parameter $\mu$. Using the approximation (\ref{3.42}) we can write
\be\n{3.44}
u\approx {\mu\over \pi}\, \ln(4\pi)+{1\over 2}\ln \left(1-{\mu\over \pi}\right)\, .
\ee
The redshift factor $u$ has maximum $u_*$
\be\n{3.44a}
u_{*}=\ln(4\pi)-{1\over 2}\left[1+\ln 2 +\ln(\ln (4\pi)) \right] \approx 1.22
\ee
at
\be\n{3.44b}
\mu=\pi(1-1/(2\ln(4\pi)))\approx 2.52\, .
\ee
For $\mu>\mu_*$ the function $u$ rapidly falls down, becoming negative
and logarithmically divergent at $\mu=\pi$.

In the same approximation we get the following expressions for the
irreducible mass $\mu_0$ and the surface gravity $\kappa$
\ba\n{3.45}
\mu_0=\mu\, \exp(-u) &\approx&
  \mu\, (4\pi)^{-{\mu\over\pi}}\, \left(1-{\mu\over\pi}\right)^{-{1\over 2}}, \\
\n{3.46}
\kappa={e^{-2u}\over 4\mu} &\approx& {1\over 4\mu}\,
(2\pi)^{2{\mu\over\pi}}\left(1-{\mu\over \pi}\right)\, .
\ea
For $\mu\to\pi$, they behave as $\mu_0\to \infty$ and $\kappa\to 0$.
Figure~\ref{u}(right) shows the irreducible mass $\mu_0$ as
a function of $\mu$.

\begin{figure}[t]
  \centerline{\epsfig{file=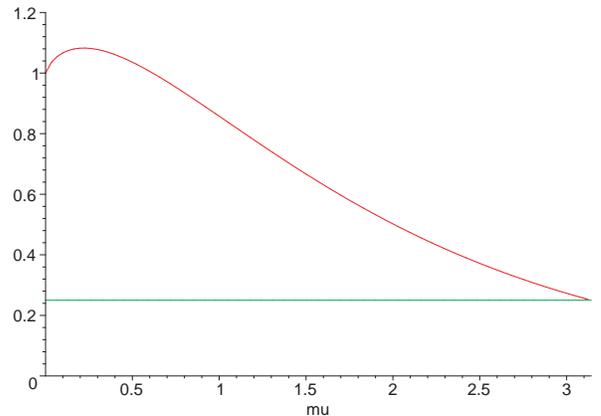, width=3in}}
  \caption{$l/(2\pi)$ as a function of $\mu$.}
  \label{polength}
\end{figure}

Another invariant characteristic of the solution is the proper distance
between the `north pole', $z=\mu$, and `south pole', $z=-\mu$, along a
geodesic connecting these poles and lying outside the black hole. This
distance, $l(\mu)$, is
\ba\n{3.45a}
l(\mu)
  &=& 2\int\limits_{\mu}^{\pi}\, dz\, e^{-U(0,z)}\\
  &\approx& 2(4\pi)^{-\mu/\pi}\, \int\limits_{\mu}^{\pi}\, dz\, \sqrt{{(z+\mu)(2\pi-z+\mu)\over (z-\mu)(2\pi-z-\mu)}}\nn
  &=&\textstyle
   2\sqrt{\pi^2-\mu^2}\, E(\varphi,k)\,
   +2\mu \sqrt{ {\pi+\mu\over \pi-\mu}}\, F(\varphi,k)\,
   -(\pi-\mu)\, ,\hspace{-1em}\nonumber
\ea
where
\be\n{3.45b}
\varphi =\sqrt{1-\mu/\pi}\, ,\hspace{1em}k={1\over \sqrt{1-(\mu/\pi)^2}}\, .
\ee
Here $F(\varphi,k)$ and $E(\varphi,k)$ are the elliptic integrals of
the first and second kind, respectively. In particular one has
\be\n{3.45c}
l(0)=2\pi \, ,\hspace{1em} l(\pi)=\pi/2\, .
\ee
Figure~\ref{polength} shows $l/(2\pi)$ as a function of $\mu$.
It might be surprising that in the limit $\mu\to\pi$, when the
coordinate distance $\Delta z$ between the poles tends to 0, the
proper distance between them remains finite. This happens because in
the same limit the surface gravity tends to 0.

\subsection{Size and shape of the event horizon}

\begin{figure*}
  \centerline{
    \epsfig{file=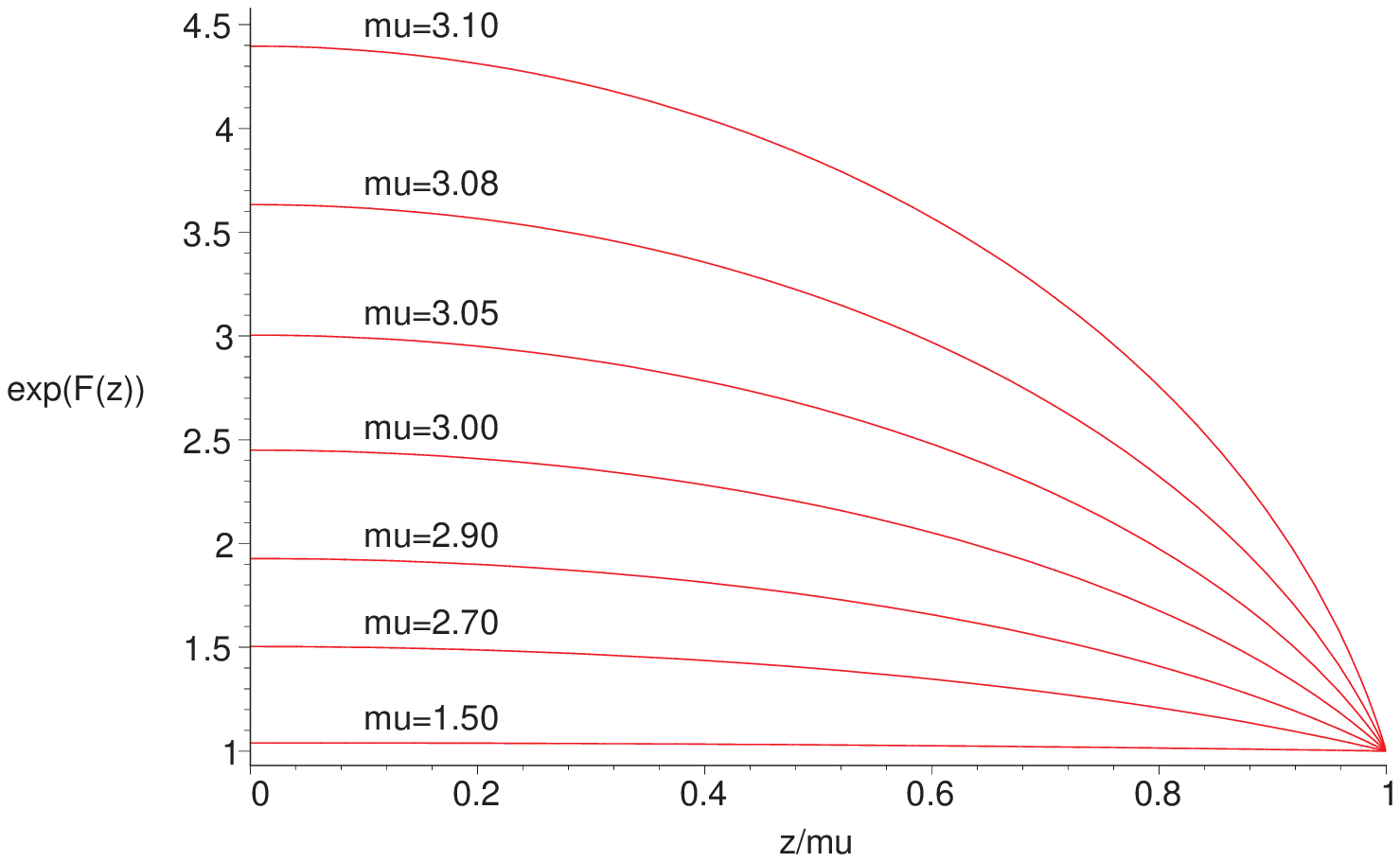, height=2in}\hspace{1cm}
    \epsfig{file=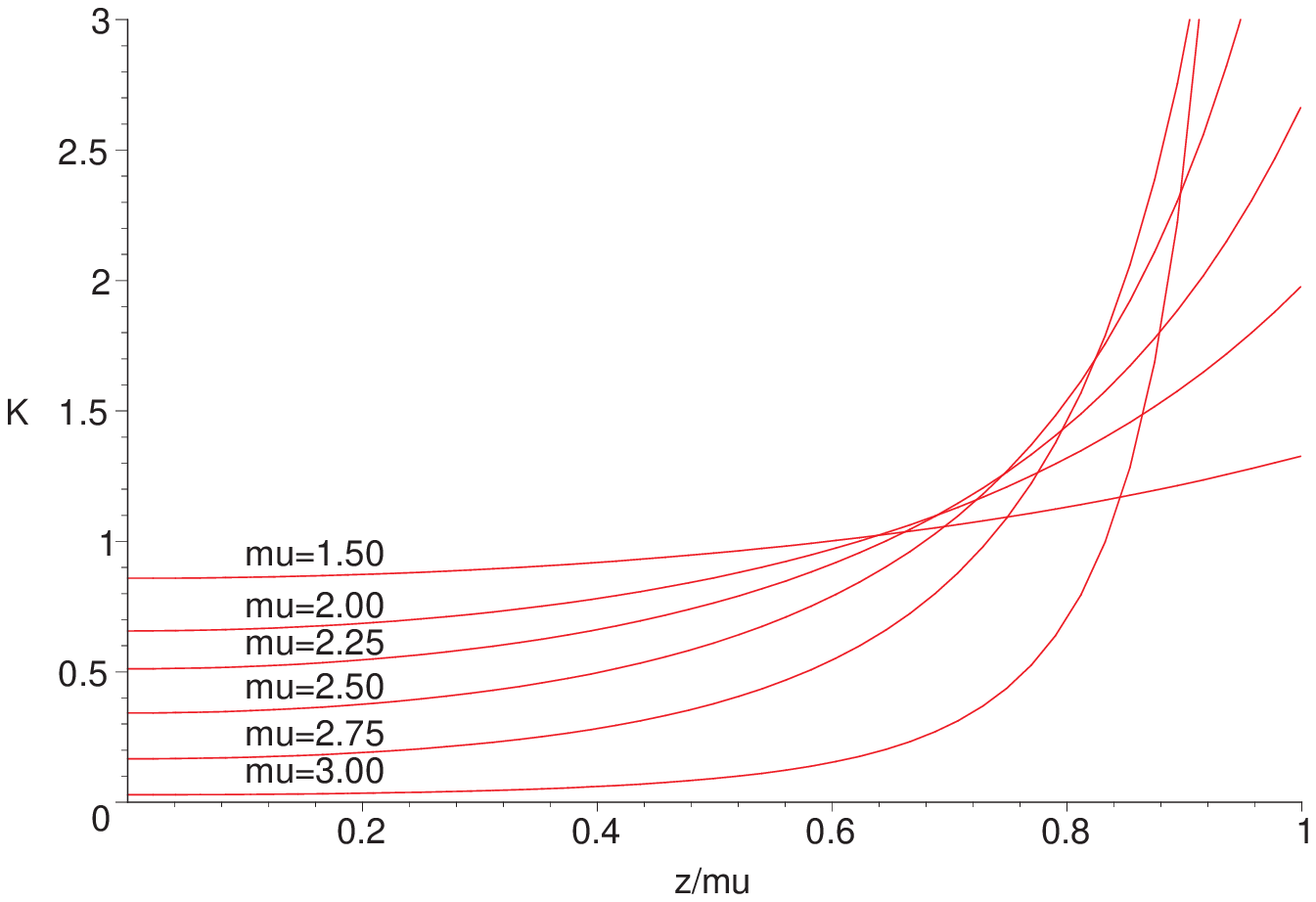, height=2in}
  }
  \caption{
    The shape function $\exp({\cal F}(z))$ (left) and the Gaussian
    curvature of the horizon $K(z)$ (right) for different values of $\mu$.
  }
  \label{shape}
  \label{gauss}
\end{figure*}

\begin{figure*}
  \medskip
  \centerline{\epsfig{file=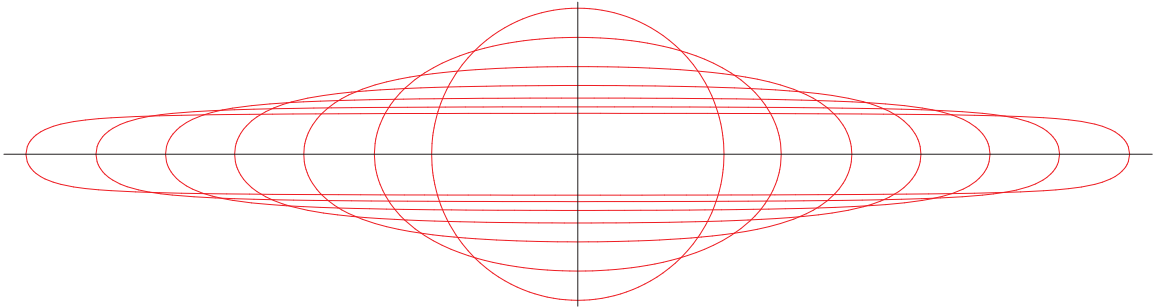, width=6in}}\medskip
  \caption{
    Embedding diagrams for the surface of the black hole horizon. By
    rotating a curve from a family shown at the plot around a horizontal
    axis one obtains surface isometric to the surface of a black hole
    described by the metric $d\sigma^2$. Different curves correspond to
    different values of $\mu$. The larger $\mu$ the more oblate is the
    form of the curve.
  }
  \label{embed}
\end{figure*}

The surface area of the distorted horizon (\ref{2.22}) written in
units $L^2$ is
\be\n{4.9}
A=16\pi \mu_0^2 \, ,
\ee
where $\mu_0$ is the irreducible mass (\ref{3.45}). The shape of the
horizon is determined by the {\em shape function}
\be\n{4.10}
{\cal F}(z)=\hat{U}(z)-u\, .
\ee
Figure~\ref{shape}(left) shows a plot of $\exp({\cal F}(z))$ for
several values of $\mu$. By multiplying the 2-metric on the horizon
$d\gamma^2$ by $(2\mu_0)^{-2}$ one obtains the metric of the 2-surface
which has the topology of a sphere $S^2$ and the surface area $4\pi$.
The metric describing this distorted sphere is
\be\n{4.11}
d\sigma^2=e^{2{\cal F}}{dz^2\over \mu^2-z^2}+e^{-2{\cal F}}
(\mu^2-z^2){d\phi^2\over \mu^2}\, .
\ee

The Gaussian curvature of the metric $d\sigma^2$ is $K={1\over 2}R$,
where $R$ is the Ricci scalar curvature. It is given by the following
expression
\be\n{4.12}
K =
e^{-2{\cal F}(z)}\, \left[1+ (\mu^2-z^2)\,[{\cal F}'' -2({\cal F}')^2
]\,
-4z {\cal F}' \right]\, .
\ee
The Gauss-Bonnet formula gives
\be\n{4.13}
\int d^2x\, \sqrt{\sigma}\, K=4\pi\, .
\ee
For the unperturbed black hole $K=1$. As a result of deformation, the
CS-black hole has $K>1$ at the poles, $z=\pm \mu$, and $K<1$ at the
`equatorial plane', $z=0$. Figure~\ref{gauss}(right), which shows $K(z)$
for different values of $\mu$, illustrates this feature. This kind of
behavior can be easily understood as a result of self-attraction of the
black hole because of the compactification of the coordinate $z$.

Using approximation (\ref{3.42}) allows one to obtain simple analytical
expressions for the shape function and the Gaussian curvature. Equations
(\ref{3.39}) and (\ref{3.43}) give
\be\n{4.14}
{\cal F}={1\over 2}\, \ln\left[{f({\mu+z\over 2}) f({\mu-z\over 2})
\over f(\mu)} \right]\approx {1\over 2}\, \ln\left[ 1+{\mu^2-z^2\over
4\pi(\pi-\mu)}\right]\, .
\ee
Let us write the metric $d\sigma^2$ in the form
\be\n{4.15}
d\sigma^2=F(z)\, dz^2+{d\phi^2\over \mu^2\, F(z)}\, ,
\ee
then in this approximation one has
\be\n{4.16}
F(z)\approx{1\over \mu^2-z^2}+{1\over 4\pi^2(1-\mu/\pi)}\, .
\ee
while the Gaussian curvature is
\be\n{4.17}
K\approx{16\pi^2(\pi-\mu)^2[ (2\pi-\mu)^2+3z^2]\over [(2\pi-\mu)^2-z^2]^3}\, .
\ee
The Gaussian curvature is positive in the interval $|z|<\mu$.

It is interesting to note that the horizon geometry of the CS-black hole
coincides (up to a constant factor) with the geometry on the 2-D
surface of the horizon of the Euclidean 4-D Kerr black hole. This fact
can be easily checked since the induced 2-D geometry of the horizon of
the Kerr black hole is (see e.g. equation 3.5.4 in \cite{FrNo:98})
\be\n{4.17a}
  dl^2=(r_+^2+a^2)\, \left[\tilde{F}(x)\, dx^2 + \frac{d\phi^2}{\tilde{F}(x)}\right],
\ee
where
\be\n{4.17b}
  \tilde{F}(x)=\frac{1}{1-x^2} - \beta^2\, ,\hspace{1em}
  \beta=\frac{a}{\sqrt{r_+^2+a^2}}\, .
\ee
Here $r_+=M+\sqrt{M^2-a^2}$ gives the position of the event horizon,
and $M$ and $a$ are the mass and the rotation parameter of the Kerr
black hole. The line element (\ref{4.15},\ref{4.16}) is obtained from
the above by coordinate redefinition $z=\mu x$ and analytic
continuation $\alpha=i\beta$, with
$\alpha = \frac{\mu}{2\pi} (1-\frac{\mu}{\pi})^{-\frac{1}{2}}$.

Denote by $l_{\text{eq}}$ the proper length of the equatorial circumference,
and by $l_{\text{pole}}$ the proper length of a closed geodesic passing
through both poles $|z|=\mu$ of the black hole horizon. Then one has
\ba\n{4.18}
l_{\text{eq}}(\mu) &\approx& 2\pi\, {\sqrt{1-\mu/\pi}\over 1-\mu/(2\pi)}\, ,\\
l_{\text{pole}}(\mu) &\approx& 4 E\left({i\mu \over 2\pi \sqrt{1-\mu/\pi}}\right)\, ,\nonumber
\ea
where $E(k)$ is the complete elliptic integral of the second kind. One
has $l_{\text{eq}}(0)=l_{\text{pole}}(0)=2\pi$ and the surface is a
round sphere. For $\mu\to \pi$ the lengths $l_{\text{eq}}\to 0$ and
$l_{\text{pole}}\to \infty$.

\subsection{Embedding diagrams for a distorted horizon}

The metric (\ref{4.15}) can be obtained as an induced geometry on a
surface of rotation $\Sigma$ embedded in a 3 dimensional Euclidean
space. Let
\be\n{6.2}
dl^2=dh^2+dr^2+r^2\, d\phi^2\,
\ee
be the metric of the Euclidean space and the surface $\Sigma$ be
determined by an equation $h=h(r)$, then the induced metric on
$\Sigma$ is
\be\n{6.3}
d\sigma^2=\left[1+\left({dh\over dr} \right)^2 \right]\, dr^2+r^2\, d\phi^2\, .
\ee
By comparing this metric with (\ref{4.15}) we get
\be\n{6.4}
r={1\over \sqrt{F(z)}}\, ,
\ee
\be\n{6.5}
\left({dh\over dz}\right)^2+\left( {dr\over dz}\right)^2= F(z)\, .
\ee
These equations imply the following differential equation for $h(z)$
\be\n{6.6}
{dh\over dz}=\sqrt{F-{{F'}^2\over 4F^3}}\, .
\ee

Figure~\ref{embed} shows the embedding diagrams for the distorted
horizon surfaces of a compactified black hole for different values of
$\mu$. The larger is the value $\mu$ the more oblate is the surface of
the horizon. For large $\mu$ close to $\pi$ it has a cigar-like
form.

\subsection{$\mu\to\pi$ limit}

Let us now discuss the properties of the spacetime in the limiting case
$\mu\to\pi$. This limit can be easily taken in the series
representation (\ref{3.34}) for the gravitational potential $U$. Since
$\sin(\pi k)=0$ for $k>0$, only the logarithmic term survives in this
limit. Thus $U(\rho,z)=\ln \rho$. Since the limiting metric is
invariant under translations in $z$-direction, it has the form of the
Kasner solution (\ref{4.4}) with $\mu=\pi$ and reads
\be\n{6.7}
ds^2=-\rho^2\, dt^2+d\rho^2+dz^2+d\phi^2\, .
\ee
This is a Rindler metric with two dimensions orthogonal to the
acceleration direction being compactified
\be\n{6.8}
z\in (-\pi,\pi)\, ,\hspace{1em}\phi \in (-\pi,\pi)\, .
\ee
Restoring the dimensionality we can write this metric as
\be\n{6.9}
dS^2=-{R^2\over L^2}\, dT^2+dR^2+dZ^2+L^2 d\phi^2\, .
\ee

\section{Discussion}\label{C}

The obtained results can be summarized as follows. If the size of a
black hole is much smaller that the size of compactification, its
distortion is small. The deformation which makes the horizon prolated
grows with the black hole mass. For large mass $\mu \ge \pi/2$ the
black hole deformation becomes profound. The pole parts of the
horizon, that is parts close to $z=-\mu$ and $z=\mu$, attract one
another. As a result of this attraction the Gaussian curvature of
regions close to black hole poles grows, while the Gaussian curvature
in the `equatorial' region falls down and the surface of the horizon
is `flattened down' in this region. For large value of the mass
$\mu$, the `flattening' effects occurs for a wide range of the
parameter $z$. Such a black hole reminds a cigar or a part of the
cylinder with two sharpened ends.

We did not include any branes in our consideration. However, we should
note that the surface $Z=0$ is a solution of the Nambu-Goto action for
a test brane. This can be easily seen, as the solution we discussed is
symmetric around the surface $Z=0$, which implies that its extrinsic
curvature vanishes there. At far distances the induced gravitational
field on the $Z=0$ submanifold is asymptotically a solution of vacuum
$(2+1)$-dimensional Einstein equations. It is not so for regions close
to the black hole. This ``violation'' of the vacuum $(2+1)$-dimensional
Einstein equations for the induced metric makes the existence of the
$(2+1)$-dimensional black hole on the brane possible.

In our work we did not find any indications on instability of a black
hole which might be interpreted as connected with the Gregory-Laflamme
instability \cite{GrLa:93,GrLa:94}. It may not be surprising since
these kind of instabilities are expected in spacetimes with higher
number of dimensions (see e.g.~\cite{GrLa:88,Kol:02,Wise:02a,Wise:02b}).

\begin{figure}[t]
  \centerline{\epsfig{file=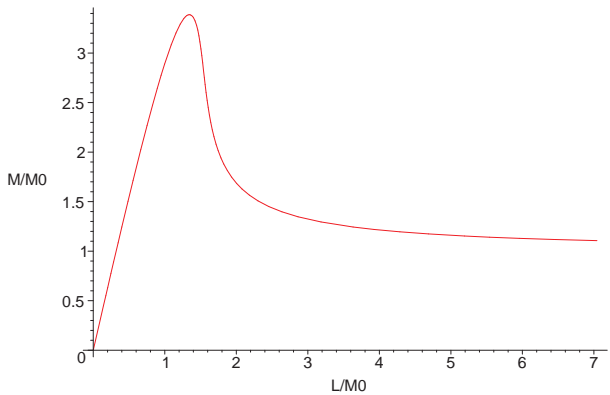, height=2in}}
  \caption{$M$ as a function of $L$ for fixed $M_0$.}
  \label{LM}
\end{figure}

On the other hand, a solution describing a black hole in a compactified
spacetime may be unstable for a different reason. The nature of this instability is the
following. In our set-up we fix a radius of compactification $L$. In a
flat spacetime we can choose parameter $L$ arbitrarily and the energy
of the system, being equal to zero, does not depend on this choice. The
situation is different in the presence of a black hole. Consider a
black hole of a given area, that is with a fixed parameter $M_0$. Since
the black hole entropy, which is proportional to the area, remains
unchanged for quasi-stationary adiabatic processes, one may consider
different states of a black hole with a given $M_0$. $L$ plays a role
of an independent parameter, specifying a solution. In particular one
has
\be\n{5.1}
  M_0={M\, (4\pi)^{-M/(\pi L)}\over \sqrt{1-M/(\pi L)}}\, .
\ee
This relation shows that for fixed $M_0$ the energy of the system $M$
depends on compactification radius $L$. The plot of the function $M(L)$
is shown at Figure~\ref{LM}. For $L=L_*=1.345 M_0$ the mass $M$ has
maximum $M=M_*=3.3877 M_0$. At the corresponding value $\mu_*=2.52$ the
function $u(\mu)$ has its maximum. Thus if one starts with a system
with $L>L_*$ then a positive variation of parameter $L$ will decrease
the energy of the system. In this case the lowest energy state
corresponds to $L\to\infty$, so that a stable solution will be an
isolated Schwarzschild black hole in an empty spacetime without any
compactifications. In the opposite case, $L<L_*$, the energy decreases
when $L\to 0$. In this limit $M\approx \pi L$ and hence it corresponds
to a limiting solution $\mu\to\pi$. The limiting metric is given by
(\ref{6.9}). The corresponding spacetime is a 2-D torus
compactification of the Rindler metric.

This argument, based on the energy consideration, indicates a possible
instability of a compactified spacetime with a black hole with respect
to compactified dimension either `unwrapping' completely or being
`swallowed' by a black hole. While `unwrapping' of the extra dimension
may be prevented by the usual stabilization mechanisms, the other
instability regime might not be so benign. It is interesting to check
whether this conjecture is correct by standard perturbation analysis.

\section*{Acknowledgments}
This work was partly supported by the Natural Sciences and Engineering
Research Council of Canada. One of the authors (V.F.) is grateful to
the Killam Trust for its financial support.




\begin{thebibliography}{9}

\bibitem{GiWi:86}
G.~W.~Gibbons and D.~L.~Wiltshire,
{\it Black holes in Kaluza-Klein theory},
Annals Phys.\  {\bf 167}, 201 (1986)
[Erratum-ibid.\  {\bf 176}, 393 (1987)].

\bibitem{Lars:00}
F.~Larsen,
{\it Kaluza-Klein black holes in string theory},
\texttt{hep-th/0002166}.

\bibitem{Myers:87}
R.~C.~Myers,
{\it Higher dimensional black holes in compactified space-times},
Phys.\ Rev.\ D {\bf 35}, 455 (1987).

\bibitem{ChHaRe:00}
A.~Chamblin, S.~W.~Hawking and H.~S.~Reall,
{\it Brane-world black holes},
Phys.\ Rev.\ D {\bf 61}, 065007 (2000)
[\texttt{hep-th/9909205}].

\bibitem{EmHoMy:00}
R.~Emparan, G.~T.~Horowitz and R.~C.~Myers,
{\it Exact description of black holes on branes},
JHEP {\bf 0001}, 007 (2000)
[\texttt{hep-th/9911043}].

\bibitem{KuTaNa:03}
H.~Kudoh, T.~Tanaka and T.~Nakamura,
{\it Small localized black holes in braneworld: Formulation and numerical method},
\texttt{gr-qc/0301089}.

\bibitem{BW}
N.~Arkani-Hamed, S.~Dimopoulos and G.~R.~Dvali,
{\it The hierarchy problem and new dimensions at a millimeter},
Phys.\ Lett.\ B {\bf 429}, 263 (1998)
[\texttt{hep-ph/9803315}].

\bibitem{BoPe:90}
A.~R.~Bogojevic and L.~Perivolaropoulos,
{\it Black holes in a periodic universe},
Mod.\ Phys.\ Lett.\ A {\bf 6}, 369 (1991).

\bibitem{HaOb:02}
T.~Harmark and N.~A.~Obers,
{\it Black holes on cylinders},
JHEP {\bf 0205}, 032 (2002)
[\texttt{hep-th/0204047}].

\bibitem{IsKh:64}
W.~Israel and K.~A.~Khan,
{\it Collinear particles and Bondi dipoles in General Relativity},
Nouvo Cimento {\bf 33}, 331 (1964).

\bibitem{KoNi:94}
D.~Korotkin and H.~Nicolai,
{\it A periodic analog of the Schwarzschild solution},
\texttt{gr-qc/9403029}.

\bibitem{GeHa:81}
R.~Geroch and J.~B.~Hartle,
{\it Distorted black holes},
J.\ Math.\ Phys.\ {\bf 23} 680 (1981).

\bibitem{FaKr:01}
S.~Fairhurst and B.~Krishnan,
{\it Distorted black holes with charge},
Int.\ J.\ Mod.\ Phys.\ D {\bf 10}, 691 (2001)
[\texttt{gr-qc/0010088}].

\bibitem{Yaza:01}
S.~S.~Yazadjiev,
{\it Distorted charged dilaton black holes},
Class.\ Quant.\ Grav.\  {\bf 18}, 2105 (2001)
[\texttt{gr-qc/0012009}].

\bibitem{EmRe:02}
R.~Emparan and H.~S.~Reall,
{\it Generalized Weyl solutions},
Phys.\ Rev.\ D {\bf 65}, 084025 (2002)
[\texttt{hep-th/0110258}].

\bibitem{ShSh:00}
T.~Shiromizu and M.~Shibata,
{\it Black holes in the brane world: Time symmetric initial data},
Phys.\ Rev.\ D {\bf 62}, 127502 (2000)
[\texttt{hep-th/0007203}].

\bibitem{SoPi:02}
E.~Sorkin and T.~Piran,
{\it Initial data for black holes and black strings in 5d},
\texttt{hep-th/0211210}.

\bibitem{FrNo:98}
V.~P.~Frolov and I.~D.~Novikov,
{\it Black hole physics: Basic concepts and new developments},
Kluwer Academic Publ.\ (1998).

\bibitem{Kasner}
E.~Kasner,
{\it Geometrical theorems on Einstein's cosmological equations},
Am.\ J.\ Math.\ {\bf 43} 217 (1921).

\bibitem{PBM}
A.~P.~Prudnikov, Yu.~A.~Brychkov, and O.~I.~Marichev,
{\it Integrals and series} v.I,
Gordon and Breach Science Publishers (1986).

\bibitem{GrLa:88}
R.~Gregory and R.~Laflamme,
{\it Hypercylindrical black holes},
Phys.\ Rev.\ D {\bf 37}, 305 (1988).

\bibitem{GrLa:93}
R.~Gregory and R.~Laflamme,
{\it Black strings and p-branes are unstable},
Phys.\ Rev.\ Lett.\  {\bf 70}, 2837 (1993)
[\texttt{hep-th/9301052}].

\bibitem{GrLa:94}
R.~Gregory and R.~Laflamme,
{\it The Instability of charged black strings and p-branes},
Nucl.\ Phys.\ B {\bf 428}, 399 (1994)
[\texttt{hep-th/9404071}].

\bibitem{Kol:02}
B.~Kol,
{\it Topology change in general relativity and the black-hole black-string transition},
\texttt{hep-th/0206220}.

\bibitem{Wise:02a}
T.~Wiseman,
{\it From black strings to black holes},
Class.\ Quant.\ Grav.\  {\bf 20}, 1177 (2003)
[\texttt{hep-th/0211028}].

\bibitem{Wise:02b}
T.~Wiseman,
{\it Static axisymmetric vacuum solutions and non-uniform black strings},
Class.\ Quant.\ Grav.\  {\bf 20}, 1137 (2003)
[\texttt{hep-th/0209051}].

\end{thebibliography}
\end{document}